\documentclass[a4paper,fleqn]{cas-dc}

\usepackage{url}
\usepackage{tabularx}
\usepackage[numbers]{natbib}
\usepackage{amsmath,amsthm}
\usepackage{mathtools}
\usepackage{pifont}
\newcommand{\Checkmark}{\ding{51}}
\newcommand{\XSolidBrush}{\ding{55}}

\def\tsc#1{\csdef{#1}{\textsc{\lowercase{#1}}\xspace}}
\tsc{WGM}
\tsc{QE}

\newtheorem{theorem}{Theorem}

\newtheorem{proposition}[theorem]{Proposition}

\begin{document}
\let\WriteBookmarks\relax
\def\floatpagepagefraction{1}
\def\textpagefraction{.001}

\shorttitle{}    

\shortauthors{}  

\title [mode = title]{Gated Rotary-Enhanced Linear Attention with Rank Modulation for Long-term Sequential Recommendation}

\affiliation[1]{organization={School of Big Data \& Software Engineering, Chongqing University},
        city={Chongqing},
        postcode={401331}, 
        country={China}}

\author[1]{Juntao Hu}
\credit{Conceptualization, Methodology, Writing – original draft, Software}

\author[1]{Wei Zhou}
\ead{zhouwei@cqu.edu.cn}
\credit{Methodology, Writing – original draft, Validation, Resources}
\cortext[1]{Corresponding author}
\cormark[1]

\author[1]{Haini Cai}
\credit{Writing – original draft, Supervision, Validation, Resources}

\author[1]{Xiao Du}
\credit{Supervision, Visualization, Validation, Software}

\author[1]{Huayi Shen}
\credit{Investigation, Software}

\author[1]{Junhao Wen}
\credit{Supervision}

\begin{abstract}
In Sequential Recommendation Systems (SRSs), Transformer models have demonstrated remarkable performance but face computational and memory cost challenges, especially when modeling long-term user behavior sequences. Due to its quadratic complexity, the dot-product attention mechanism in Transformers becomes expensive for processing long sequences. By approximating the dot-product attention using elaborate mapping functions, linear attention provides a more efficient option with linear complexity. However, existing linear attention methods face three limitations: 1) they often use learnable position encodings, which incur extra computational costs in long-term sequence scenarios, 2) limited by the low-rank deficiency, they may not sufficiently account for user's fine-grained local preferences (short-lived burst of interest), and 3) they try to capture some temporary activities, but often confuse these with stable and long-term interests. This can result in unclear or less effective recommendations.
To remedy these drawbacks, we propose a long-term sequential \underline{\textbf{Rec}}ommendation model with \underline{\textbf{G}}ated \underline{\textbf{R}}otary \underline{\textbf{E}}nhanced \underline{\textbf{L}}inear \underline{\textbf{A}}ttention (\textbf{RecGRELA}). Specifically, we first propose a Rotary-Enhanced Linear Attention (RELA) module to efficiently model long-range dependency within the user's historical information using rotary position encodings. Then, to address the low-rank deficiency of linear attention, we introduce an Adaptive Rank Modulator. It incorporates a rank augmentation branch to explicitly inject local token mixing and a Gated Rank Selector to dynamically balance stable long-term preferences and transient short-term interests. Experimental results on four public benchmark datasets show that our RecGRELA achieves state-of-the-art performance compared with existing SRSs based on Recurrent Neural Networks, Transformer, and Mamba while keeping low memory overhead. Our source code is available at \url{https://github.com/TBI805/RecGRELA}.
\end{abstract}




\begin{keywords}
 \sep Sequential Recommendation \sep Linear Attention \sep Rotary Position Encoding \sep Gated Mechanism \sep Low-rank Augmentation
\end{keywords}

\maketitle


\section{Introduction}

\begin{figure*}
    \centering
    \includegraphics[width=0.9\linewidth]{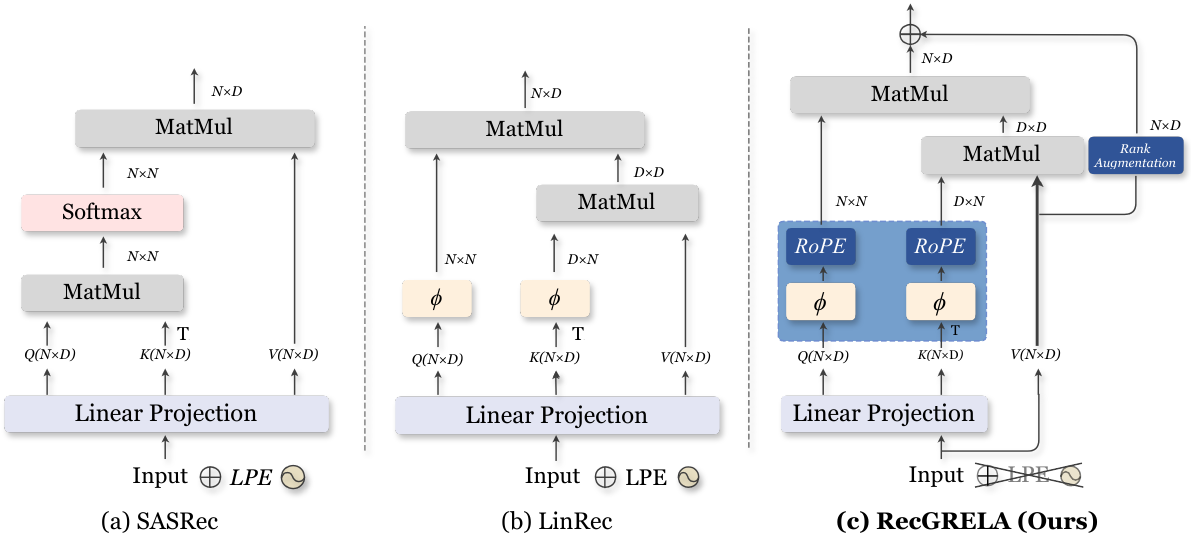}
    \caption{Illustration of differences of SASRec, LinRec, and our RecGRELA. $\phi$ denotes the kernel function.}
    \label{fig:method_comparsion}
  \end{figure*}

Sequential Recommender Systems (SRSs) \cite{ijcai2019p883,fang2020deep,pan2026survey,shen2025spatio,wang2025future} have become a fundamental technology in personalized services, significantly transforming how online platforms predict user behavior. By analyzing sequences of user interactions across e-commerce, streaming services, and social media, SRSs capture both short-term interests and long-term preference trends. However, modeling long-term dependencies is challenging, as efficiently handling long sequences demands a balance between computational cost and model performance, while also capturing the dynamic and complex patterns in user interests.

To address the challenge of modeling user behavior sequences, researchers explore various methodologies. Traditional approaches, such as those based on Markov Chains \cite{he2017translation,rendle2010factorizing} or Recurrent Neural Networks (RNNs) \cite{chen2018sequential,cui2018mv,donkers2017sequential,xu2019recurrent}, lay the groundwork but also face limitations in capturing long-range dependencies or suffered from computational inefficiencies during training and inference, particularly with increasing sequence lengths. Consequently, the field moves towards more advanced models that can handle complex sequential dependencies more effectively. This also presents new challenges.

Specifically, the first challenge is \textbf{\textit{Efficiency}}. Transformer models \cite{kang2018self,sun2019bert4rec,ijcai2019p600} emerge as a compelling replacement for earlier approaches like RNNs, leveraging their ability to process sequential data and capture both local and global preferences of users. However, the dot-product attention mechanism in Transformers exhibits quadratic complexity relative to sequence length, resulting in high costs for long sequences. To address this issue, linear attention \cite{katharopoulos2020transformers,liu2023linrec} has been proposed as an efficient alternative, providing linear complexity while maintaining the modeling capabilities. Position encoding is widely adopted to capture the sequential order of items users interact with due to its ability to enhance the positional awareness of contextual vectors. As illustrated in Fig. \ref{fig:method_comparsion}(a) and (b), the above attention-based methods rely on learnable position encoding. The learned positional encoding requires more parameters than fixed positional encoding schemes since each position needs its own learned vector, and they do not have any inherent structure that captures the notion of relative distances between positions. To handle these drawbacks, some works \cite{su2024roformer,tian2024eulerformer} employ an improved Rotary Position Encoding (RoPE), which injects relative positional information in a more efficient manner. However, they still use the dot-product attention mechanism, which limits its scalability for long sequences. Furthermore, the RoPE exhibits inefficiency in multi-head attention scenarios, restricting the model's overall expressive capability. 

Linear attention, while effective for grasping overarching long-term dependencies, can struggle to identify \textbf{\textit{Fine-Grained Local Preferences}}. Limited by the low-rank deficiency, the attention matrix in linear attention is restricted to a low rank, which limits the diversity of token-to-token interactions. In recommendation contexts, user interactions often reveal dense, localized activity sequences. For instance, a user might intensely engage with a specific series of related items. They could watch multiple episodes of \textit{The Avengers} and then immediately consume supplementary content detailing Marvel's production process. Subsequently, they might delve into documentaries that explore the creators behind such cinematic universes. These closely related, short-term sequences of engagement are crucial for understanding immediate user intent, yet many existing approaches fail to capture their significance in long-term sequential recommendation due to this capacity bottleneck.

Even when linear attention successfully captures both long-term interests and short-term local preferences, a significant challenge persists: \textbf{\textit{when a user's preferences really change, it's hard for the model to identify them}}. For instance, consider a user like historical documentaries over a weekend, binge-watches several interconnected Marvel movies like \textit{The Avengers} series, and then explores some behind-the-scenes content about their production. A model that struggles to differentiate signals might misinterpret this intense but potentially short-lived burst of activity as a long-term shift in preference. Consequently, it might aggressively recommend a deluge of Marvel-related content for weeks – from other superhero films they've already seen, to comic book adaptations, or even tangentially related action movies. This happens because the model overweights fleeting interests. The user might simply want a weekend of superhero fun and is now looking for a historical drama, getting frustrated by the repetitive and misaligned suggestions.

To solve the aforementioned challenges, we propose \textbf{RecGRELA}, a novel long-term sequential \textbf{\underline{Rec}}ommendation model with \textbf{\underline{G}}ated \textbf{\underline{R}}otary \textbf{\underline{E}}nhanced \textbf{\underline{L}}inear \underline{A}ttention block (as shown in Fig. \ref{fig:method_comparsion}(c)) to effectively capture long-range dependency. Specifically, we design a Rotary Enhanced Linear Attention (RELA) module. RELA builds upon the efficiency of linear attention, directly addressing the quadratic complexity issue of standard Transformers, but significantly enhances its positional awareness by incorporating Rotary Position Encoding (RoPE). This allows the model to effectively capture relative inter-item positional relationships in a parameter-efficient manner, overcoming the drawbacks of purely learnable or fixed encodings and the inefficiency of RoPE in standard multi-head dot-product attention. Concurrently, to address the low-rank deficiency of linear attention, we introduce an Adaptive Rank Modulator (ARM). It first leverages a Rank Augmentation Branch to explicitly inject short-range transition patterns via causal 1D convolution, providing theoretical insights based on matrix rank analysis. We further design a Gated Rank Selector to dynamically modulate the information flow, distinguishing true shifts in long-term user interests from temporary bursts of local activities. With this gated mechanism, we remove the linear projection on the value matrix of the attention module to reduce computational cost.

The contributions of this work are summarized as follows: 
\begin{itemize} \item We propose a Rotary Enhanced Linear Attention (RELA) module that integrates Rotary Position Encoding (RoPE) into the linear attention mechanism to address efficiency issues in long-sequence modeling. By replacing dot-product attention with linear attention, our approach reduces computational costs while maintaining the ability to capture inter-item positional relationships. 
\item We propose an Adaptive Rank Modulator (ARM) to overcome the low-rank deficiency of linear attention. It features a Rank Augmentation Branch to explicitly model dense local interactions without sacrificing computational efficiency. 
\item We introduce a Gated Rank Selector within ARM to dynamically modulate the information flow, allowing the model to distinguish true shifts in user preferences from transient local interests. Based on these designs, we further construct the RecGRELA, a novel long-term sequential recommendation paradigm that achieves competitive performance compared with state-of-the-art SRSs on four benchmark datasets, keeping low memory overhead.
\end{itemize}

\section{PRELIMINARY}
\subsection{Problem Definition}

Considering a set of users $\mathcal{U}=\{u_i\}_{i=1}^{|\mathcal{U}|}$ and a set of items $\mathcal{V}=\{v_{i}\}_{i=1}^{|\mathcal{V}|}$. We use $\mathcal{S}_u=\left\{v_t\right\}_{t=1}^T$ to denote an interaction sequence of $T$ chronologically ordered items that the user $u$ has interacted with. Given a user interaction sequence $\mathcal{S}_u$, the task of sequential recommendation is to compute a ranking list consisting of Top-$k$ items that the user $u$ is likely to interact with at the next time step $T + 1$. We define $N$ as the sequential length and $D$ as its corresponding dimension. Typically, the $N$ is often relatively short, but in some complex scenarios, where the $N$ satisfies $N$ > 1.5$D$ (e.g., when $D$=64, the $N$ > 96), the sequence is regarded as ``\textbf{long-term}'' \cite{liu2023linrec}.


\subsection{Dot-Product Attention and Linear Attention}
Let $\boldsymbol{X} \in \mathbb{R}^{N \times D}$ denotes an input sequence. Formally, the input sequence $\boldsymbol{X}$ is projected by three projection matrices $\boldsymbol{W}_Q$, $\boldsymbol{W}_K, \boldsymbol{W}_V\in \mathbb{R}^{D \times d}$, to corresponding representations $\boldsymbol{Q}=\boldsymbol{X}\boldsymbol{W}_Q$, $\boldsymbol{K}=\boldsymbol{X}\boldsymbol{W}_K$, and $\boldsymbol{V}=\boldsymbol{X}\boldsymbol{W}_V$. The \textbf{dot-product attention} \cite{vaswani2017attention} can be written as:
\begin{equation}
    \quad {\boldsymbol{O}}_m=\sum_{n=1}^N \frac{\exp \left(\boldsymbol{Q}_m\boldsymbol{K}_n^{\top} / \sqrt{d}\right)}{\sum_{n=1}^N\exp \left(\boldsymbol{Q}_m \boldsymbol{K}_n^{\top} / \sqrt{d}\right)} \boldsymbol{V}_n,
    \label{eq1}
\end{equation}
where $\boldsymbol{Q}_m, \boldsymbol{K}_n, \boldsymbol{V}_n \in \mathbb{R}^{1\times d}$, and $d$ is head dimension. In this case, calculating attention weights in dot-product attention leads to the computation complexity of $\mathcal{O}(N^2)$ with respect to the input sequence length $N$. For autoregressive modeling, dot-product attention can be implemented by incorporating a masking mechanism, which restricts the influence on the $m$-th position to a position $n$ satisfying $n \leq m$.

\textbf{Linear attention} \cite{katharopoulos2020transformers} is proposed as an effective alternative that reduces the computation complexity from $\mathcal{O}(N^2)$ to $\mathcal{O}(N)$ compared with dot-product attention. Specifically, linear attention modifies the dot-product attention by replacing the non-linear Softmax operation with linear normalization, while incorporating an additional kernel function $\phi$ in query matrix $\boldsymbol{Q}$ and key matrix $\boldsymbol{K}$:
\begin{equation}
\boldsymbol{O}_{m}=\sum_{n=1}^{N}\frac{\phi\left(\boldsymbol{Q}_{m}\right)\phi\left(\boldsymbol{K}_{n}\right)^\top}{\sum_{n=1}^{N}\phi\left(\boldsymbol{Q}_{m}\right)\phi\left(\boldsymbol{K}_{n}\right)^\top}\boldsymbol{V}_{n}.
\end{equation}
By utilizing the associative property of matrix multiplication, the computation is reordered from $(\boldsymbol{Q}\boldsymbol{K}^\top)\boldsymbol{V}$ to $\boldsymbol{Q}(\boldsymbol{K}^{\top}\boldsymbol{V})$:
\begin{equation}
    \boldsymbol{O}_m=\frac{\phi\left(\boldsymbol{Q}_m\right)\left(\sum_{n=1}^N\phi\left(\boldsymbol{K}_n\right)^\top\boldsymbol{V}_n\right)}{\phi(\boldsymbol{Q}_m)\left(\sum_{n=1}^N\phi\left(\boldsymbol{K}_n\right)^\top\right)},
    \label{EQ3}
\end{equation}
where the computational complexity is decreased to $\mathcal{O}(N)$. In \cite{katharopoulos2020transformers}, the kernel function employs the Exponential Linear Unit (ELU) activation \cite{clevert2015fast}, which prevents gradient vanishing when processing negative values of $\boldsymbol{X}$. In our work, we follow this setting for a fair comparison.

\subsection{Position Encoding}
\textbf{Absolute Position Encoding} (\textbf{APE}) \cite{vaswani2017attention} incorporates fixed values into the embeddings of sequence elements, establishing a consistent reference frame across all tokens. For a token at the position $t$ within a sequence of length $N$, this encoding is computed through a deterministic function utilizing paired sine and cosine operations across different frequencies. The formulation is denoted as follows:
\begin{equation}
    \begin{aligned}
    PE\left( p, 2t \right) &=\sin \left( p/10000^{2t/\left( D /2 \right)} \right),\\
    PE\left( p, 2t+1 \right) &=\cos \left( p/10000^{2t/\left( D /2 \right)} \right),
    \end{aligned}
    \label{EQ4}
\end{equation}
where $\emph{p}$ is the position of the item, $\emph{t}$ is the dimension of $\emph{p}$.

\textbf{Learnable Position Encoding} (\textbf{LPE}) \cite{wang2020position, kang2018self} adopts a distinct methodology by modeling positional information as learnable parameters that are optimized during the training process. Specifically, for an item embedding \textbf{\textit{x}} at position $m$, the transformation is performed as follows:
\begin{equation}
    \textbf{\textit{x}}_m = \textbf{\textit{x}} + p_m,
\end{equation}
where $p_i$ denotes the APE or LPE.

\textbf{Rotary Position Encoding} (\textbf{RoPE}) \cite{su2024roformer} captures
relative positional information \cite{shaw2018self} through absolute position encoding methods via encoding the relative positional information with a rotation matrix $\mathbf{R}\in\mathbb{C}^{N\times(d/2)}$:
\begin{equation}
    \mathbf{R}\left(m,t\right)=e^{i\theta_tm},
\end{equation}
where $i$ denotes the imaginary unit in complex numbers. Following \cite{vaswani2017attention}, the $\theta_t$ is set to $10000^{-t/(d/2)}$, where $t\in[1,2,...,d/2]$. Then RoPE converts $\boldsymbol{Q}_m$, $\boldsymbol{K}_n\in\mathbb{R}^{N\times{d}}$ to complex vector $\boldsymbol{\bar{Q}}_m$, $\boldsymbol{\bar{K}}_n\in\mathbb{C}^{N\times(d/2)}$ by considering (2$t$)-th dim as real part
and (2$t$ + 1)-th dim as imaginary part. The RoPE is applied to complex vectors of query and key as:
\begin{equation}
    \begin{aligned}
    \bar{\boldsymbol{Q}}^{\prime}_m&=\boldsymbol{\bar{Q}}_m\odot\mathbf{R}\left(m,t\right),\\
    \bar{\boldsymbol{K}}^{\prime}_n&=\boldsymbol{\bar{K}}_n\odot\mathbf{R}\left(n,t\right),\\ 
    \boldsymbol{A}_{(m,n)}^{\prime}&=\mathrm{Re}[\bar{\boldsymbol{Q}}^{\prime}_m\bar{\boldsymbol{K}}^{\prime}_n]=\mathrm{Re}[\boldsymbol{\bar{Q}}_m\boldsymbol{\bar{K}}_n^*e^{i(m-n)\theta_t}],
    \end{aligned}
    \label{EQ7}
\end{equation}
where $\mathrm{Re}[\cdot]$ represents real part of the complex vector and * means complex conjugates, and $\odot$ denotes Hadamard product. Note that the attention matrix $\boldsymbol{A}_{(m,n)}$ with RoPE implies relative position in rotation form $e^{i(m-n)\theta_t}$ for $(D/2)$ frequencies. We can define the RoPE operation $\mathcal{R}\left(\cdot\right)$ to an input real vector $\boldsymbol{X}$ as follows:
\begin{equation}
    \mathcal{R}\left(\boldsymbol{X}\right)=\mathcal{C}^{-1}\left(\mathcal{C}\left(\boldsymbol{X}\right)\odot\mathbf{R}\right),
\end{equation}
where $\mathcal{C}\left(\cdot\right)$ converts a real-valued tensor into a complex-valued tensor, and $\mathcal{C}^{-1}\left(\cdot\right)$ converts a complex-valued tensor into a real-valued tensor.

\section{METHODOLOGY}

\begin{figure*}
    \centering
    \includegraphics[width=0.95\textwidth]{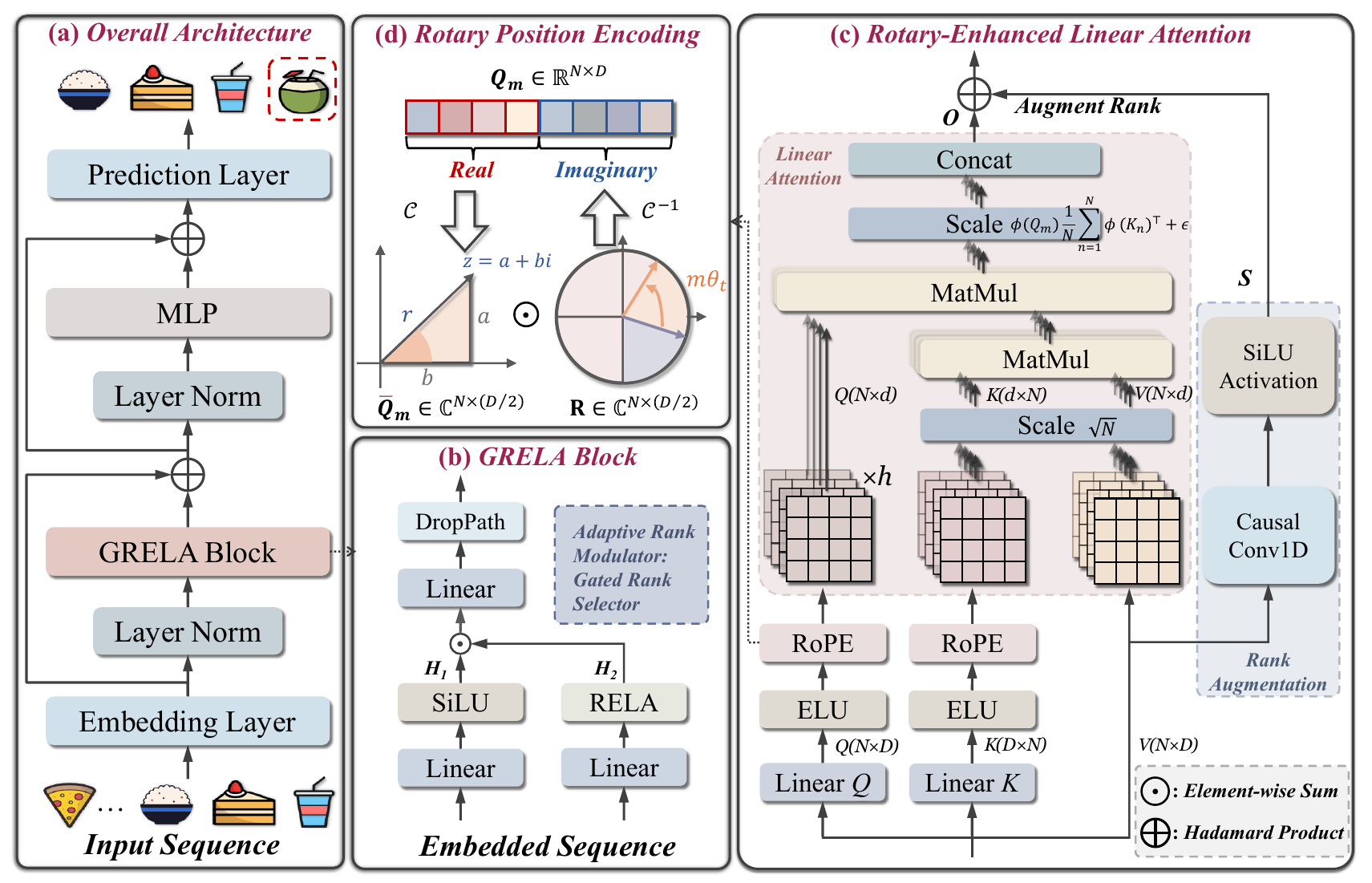}	
    \caption{The overview of our proposed architecture. The overall architecture (a) processes an input sequence through an Embedding Layer, which is then fed into a stack of $L$ identical encoder layers. Each encoder layer consists of a GRELA Block and an MLP, finally leading to a Prediction Layer. The central component is the GRELA Block (b), which involves the RELA module (c). Within the RELA module, query ($Q$) and key ($K$) are processed by ELU activation and the RoPE (d). Attention scores are computed via matrix multiplication of the RoPE-enhanced $Q$ and $K$ (after $K$ is scaled), followed by another scaling and a concatenation step. To overcome the low-rank deficiency of linear attention and capture fine-grained user preferences, an Adaptive Rank Modulator (ARM) leverages a Rank Augmentation Branch to explicitly inject local context into the value ($V$) projection. Finally, a Gated Rank Selector (GRS) dynamically modulates the information flow between global and local interactions.}
    \label{fig:overall}
\end{figure*}

In this section, we introduce our proposed framework, RecGRELA. The framework overview is shown in Figure \ref{fig:overall}. The presentation is structured as follows: initially, we introduce the embedding layer; secondly, we elaborate on the core components of the GRELA Block; subsequently, we present the details of the Multi-Layer Perceptron, and finally, we describe the prediction layer and training strategies. 


\subsection{Embedding Layer}
Following \cite{kang2018self}, we utilize an embedding layer to map the training sequence $\mathcal{S}_u$ into a high-dimensional space. The embedding layer uses a learnable embedding matrix $\boldsymbol{E} \in \mathbb{R}^{|\mathcal{V}| \times D}$, where $\mathcal{V}$ denotes the item set and $D$ represents the embedding dimension. Given an input interaction sequence $\mathcal{S}_u$ of length $N$, it can be projected into the initial embedding matrix $\boldsymbol{H}_o \in \mathbb{R}^{N \times D}$ as follows:
\begin{equation}
    \boldsymbol{H}_o=\boldsymbol{E}\mathcal{S}_u.
\end{equation}
Then we apply the Dropout \cite{srivastava2014dropout} and layer normalization \cite{lei2016layer} operations to obtain the final embedding $\boldsymbol{H} \in \mathbb{R}^{N \times D}$:
\begin{equation}
    \boldsymbol{H}=\operatorname{LayerNorm}\left(\operatorname{Dropout}\left(\boldsymbol{H}_o\right)\right).
\end{equation}
Attention-based sequential recommendation models \cite{kang2018self,sun2019bert4rec,li2020time,liu2023linrec} typically inject a learnable position encoding into the final embedding, while our method captures position awareness with rotary position encoding by multiplying the rotation matrix with outputs of the kernel function in linear attention. This design eliminates the need to incorporate positional information into the embedding layers.

\subsection{GRELA Block}
In this part, we present the design of our proposed GRELA Block, which includes the Rotary-Enhanced Linear Attention module and the Adaptive Rank Modulator.

\subsubsection{Rotary-Enhanced Linear Attention}
\textbf{Overall Design.} Previous work \cite{liu2023linrec} provides $\mathcal{O}(N)$ computational complexity of attention mechanism to alleviate the issue of quadratic complexity in dot-product attention for sequential recommendation. However, it ignores the high cost of learnable position encoding in long-sequence modeling. Meanwhile, some works \cite{tian2024eulerformer,lopez2024positional} notice these issues and propose improved RoPE to address them, but unfortunately, they fail to address the problem of dot-product attention with $\mathcal{O}(N^2)$ complexity. To bridge this gap, we propose a novel Rotary-Enhanced Linear Attention (RELA) mechanism, aiming to simultaneously resolve the issues of quadratic complexity and ineffective position encoding in long-sequence modeling. The motivation stems from the observation that other position encoding methods operate directly on the input sequence or the attention matrix. However, linear attention does not compute the attention matrix, which means other methods can only modify the input sequence. In contrast, RoPE achieves relative position encoding by leveraging absolute position encoding, by applying transformations only to the queries $\boldsymbol{Q}$ and keys $\boldsymbol{K}$. This design makes RoPE suitable for integration with linear attention. In a general situation, following Eq.(\ref{EQ7}), we revise Eq. (\ref{EQ3}) as follows:

\begin{equation}
    \boldsymbol{O}_m=\frac{\mathcal{R}\left(\phi\left(\boldsymbol{Q}_m\right)\right)\left(\mathcal{R}\left(\sum_{n=1}^N\phi\left(\boldsymbol{K}_n\right)^{\top}\right)\boldsymbol{V}_n\right)}{\phi(\boldsymbol{Q}_m)\left(\sum_{n=1}^N\phi\left(\boldsymbol{K}_n\right)^\top\right)},
\end{equation}
where $\mathcal{R}\left(\cdot\right)$ denotes injecting the RoPE into queries or keys. Following \cite{katharopoulos2020transformers}, we set $\phi(x) = \operatorname{ELU}(x) + 1$. This kernel function ensures that the denominator remains positive to avoid the risk of division by zero, while the summation in the numerator also contains no negative terms. 

Inspired by techniques for stabilizing gradients in dot-product attention \cite{vaswani2017attention}, we further incorporate scaling factors. Specifically, the keys $\boldsymbol{K}$ and values $\boldsymbol{V}$ are scaled by $\sqrt{N}$ (where $N$ denotes the sequence length). This emulates the scaling mechanism used in conventional dot-product attention and helps regulate the dynamic range of the resulting key-value memory. Furthermore, the normalization term in the denominator, which is critical for linear attention, is constructed using the dot product between the query $\phi(\boldsymbol{Q}_m)$ and the mean of keys $\phi(\boldsymbol{K}_n)$. A small constant $\epsilon$ is added to this denominator for numerical stability, preventing division by zero and ensuring well-behaved gradients. This approach enables more fine-grained control over the attention distribution. The final RELA is formulated as follows:

\begin{equation}
    \boldsymbol{O}_m=\frac{\mathcal{R}\left(\phi\left(\boldsymbol{Q}_m\right)\right)\left(\left(\mathcal{R}\left(\sum_{n=1}^N\phi\left(\boldsymbol{K}_n \right)^{\top}\right)\boldsymbol{V}_n/\sqrt{N}\right)\right)}{\phi(\boldsymbol{Q}_m)\frac{1}{N}\sum_{n=1}^N\phi\left(\boldsymbol{K}_n\right)^{\top} + \epsilon}.
    \label{eq:RELA}
\end{equation}

\textbf{Theoretical Justification: Translation Equivariance.}
We now show that incorporating RoPE into the kernelized query/key yields an interaction that depends only on relative offsets, which is the desired translation-equivariant inductive bias for sequential modeling.

\begin{proposition}[Relative Position Dependence / Translation Equivariance]
\label{prop:translation}
Let $\mathcal{R}_m: \mathbb{R}^{d} \to \mathbb{R}^{d}$ denote the RoPE operator at position $m$ (applied to each 2D feature pair with angle $m\theta_j$ for frequency $\theta_j$).
For any kernelized query $\boldsymbol{q}=\phi(\boldsymbol{Q}_m)\in \mathbb{R}^{d}$ and key $\boldsymbol{k}=\phi(\boldsymbol{K}_n)\in \mathbb{R}^{d}$, define the RoPE interaction
\begin{equation}
s(m,n;\boldsymbol{q},\boldsymbol{k}) \;=\; \langle \mathcal{R}_m(\boldsymbol{q}), \mathcal{R}_n(\boldsymbol{k}) \rangle .
\end{equation}
Then the interaction is \emph{translation equivariant} in the sense that for any integer shift $t$,
\begin{equation}
s(m+t,n+t;\boldsymbol{q},\boldsymbol{k}) \;=\; s(m,n;\boldsymbol{q},\boldsymbol{k}),
\end{equation}
and equivalently $s(m,n;\boldsymbol{q},\boldsymbol{k})$ depends on $(m-n)$ only (given $\boldsymbol{q},\boldsymbol{k}$).
\end{proposition}

\begin{proof}
It suffices to verify the claim on each 2D feature pair.
For the $j$-th pair, write the real 2D components as a complex number
$q_j = q_{j,1} + i q_{j,2}$ and $k_j = k_{j,1} + i k_{j,2}$.
RoPE applies a unit-modulus rotation:
\begin{equation}
\mathcal{R}_m(q_j) = q_j e^{i m\theta_j}, \qquad \mathcal{R}_n(k_j) = k_j e^{i n\theta_j}.
\end{equation}
Using the Hermitian inner product for the complex form, the contribution of this pair to the dot product can be written as
\begin{equation}
    \begin{aligned}
    \langle \mathcal{R}_m(q_j), \mathcal{R}_n(k_j) \rangle
    &= \operatorname{Re}\!\left( (\mathcal{R}_m(q_j))^* \, \mathcal{R}_n(k_j) \right) \\
    &= \operatorname{Re}\!\left( q_j^* k_j \, e^{i(n-m)\theta_j} \right).
    \end{aligned}
\end{equation}
Crucially, the positional factor appears only through the relative offset $(n-m)$.
Therefore, for any shift $t$,
\begin{equation}
\operatorname{Re}\!\left( q_j^* k_j \, e^{i((n+t)-(m+t))\theta_j} \right)
= \operatorname{Re}\!\left( q_j^* k_j \, e^{i(n-m)\theta_j} \right),
\end{equation}
which proves translation equivariance for this pair.
Summing over all pairs $j$ yields the full dot product $s(m,n;\boldsymbol{q},\boldsymbol{k})$ and preserves the same invariance.
\end{proof}

\noindent\textbf{Remark.}
The above result does not require the interaction to be explicitly factorized into separate semantic and positional components.
Instead, it guarantees that RoPE injects position information exclusively through the relative displacement $(m-n)$, while the content-dependent term is fully contained in $q_j^*k_j$.

\textbf{Implication.} This result guarantees that the attention mechanism is \textit{translation equivariant}: shifting the entire sequence by $\Delta$ positions does not change the pairwise attention weights. This is a key inductive bias missing in standard Linear Attention, which treats context as a bag-of-features without temporal ordering. The detailed derivation of RoPE follows standard literature \cite{su2024roformer}.
\begin{equation}
    \boldsymbol{Q}_m=\left[\boldsymbol{Q}_m^{(1)}, \boldsymbol{Q}_m^{(2)}\right]=\left[\boldsymbol{Q}_m^{(1)}+i \boldsymbol{Q}_m^{(2)}\right],
\end{equation}
where $\boldsymbol{Q}_m^{(1)}$ and $\boldsymbol{Q}_m^{(2)}$ indicate the components of $\boldsymbol{Q}_m$ expressed in the 2D coordinates. Based on Euler's formula, $e^{im\theta}$ can be rewritten as follows:
\begin{equation}
    e^{im\theta} = \cos m\theta + i\sin m\theta.
\end{equation}
Then, we have:
\begin{small}
\begin{equation}
    \begin{aligned}
        \boldsymbol{Q}_m e^{i m \theta} & =\left(\boldsymbol{Q}_m^{(1)}+i \boldsymbol{Q}_m^{(2)}\right)(\cos (m \theta)+i \sin (m \theta)) \\
        =\left(\boldsymbol{Q}_m^{(1)} \cos (m \theta)\right. & \left.-\boldsymbol{Q}_m^{(2)} \sin (m \theta)\right)+i\left(\boldsymbol{Q}_m^{(2)} \cos (m \theta)+\boldsymbol{Q}_m^{(1)} \sin (m \theta)\right).
    \end{aligned}
\end{equation}
\end{small}
Then, the result is re-expressed in the form of a real-valued vector:
\begin{equation}
    \begin{aligned}
        \boldsymbol{Q}_m e^{i m \theta} = [&\boldsymbol{Q}_m^{(1)}\cos(m\theta)-\boldsymbol{Q}_m^{(2)}\sin(m\theta), \\&\boldsymbol{Q}_m^{(2)}\cos(m\theta)+\boldsymbol{Q}_m^{(1)}\sin(m\theta)] \\
        & = \left(\begin{array}{cc}
        \cos m \theta & -\sin m \theta \\
        \sin m \theta & \cos m \theta
        \end{array}\right)\binom{\boldsymbol{Q}_m^{(1)}}{\boldsymbol{Q}_m^{(2)}}.
    \end{aligned}
\end{equation}
Evidently, the matrix $\left(\begin{array}{cc}\cos m \theta & -\sin m \theta \\ \sin m \theta & \cos m \theta \end{array}\right)$ describes a rotational transformation applied to the embedded sequence.

\subsubsection{Adaptive Rank Modulator}
\label{sec:arm}
While linear attention optimizes computational efficiency, it suffers from a  expressiveness bottleneck known as the \textit{low-rank deficiency} \cite{han2023flatten}. Since the attention mixing is constrained by the feature dimension $d$ (where $d \ll N$), the model may miss fine-grained local interactions. To address this, we propose the \textbf{Adaptive Rank Modulator (ARM)}, which integrates a \textit{Rank Augmentation Branch} to inject explicit local token mixing and a gated mechanism for dynamic selection.

\textbf{Rank Augmentation Branch.}
ARM adds a lightweight local pathway in parallel to linear attention. Concretely, we apply a learnable \emph{causal 1D convolution} over the value sequence $\boldsymbol{V}$ to extract short-range transition patterns, followed by a nonlinearity:
\begin{equation}
    \boldsymbol{S} = \sigma\left( \operatorname{CConv}_{\boldsymbol{\Theta}}(\boldsymbol{V}) \right),
\end{equation}
where $\operatorname{CConv}_{\boldsymbol{\Theta}}(\cdot)$ denotes the causal convolution operator with learnable kernel parameters $\boldsymbol{\Theta}$, and $\sigma(\cdot)$ is the SiLU activation.
In practice, this branch introduces explicit neighborhood token mixing that complements the global mixing produced by linear attention.

\textbf{Why this helps.}
Although the outputs $\boldsymbol{O}\in\mathbb{R}^{N\times d}$ and $\boldsymbol{S}\in\mathbb{R}^{N\times d}$ are both bounded by $d$ in matrix rank due to their shapes, the key limitation of linear attention comes from how tokens are mixed across positions.
Ignoring normalization for clarity, linear attention induces a token-mixing matrix
\begin{equation}
    \boldsymbol{A} \;=\; \phi(\boldsymbol{Q})\,\phi(\boldsymbol{K})^{\top}\in\mathbb{R}^{N\times N},
    \qquad \operatorname{rank}(\boldsymbol{A})\le d\ll N,
\end{equation}
which limits the diversity of token-to-token interactions.
The causal convolution branch provides an additional local mixing path.
Thus, the fused representation can be written as
\begin{equation}
    \boldsymbol{O}_f \;=\; \boldsymbol{O}+\boldsymbol{S}.
    \label{EQ17}
\end{equation}
This simple additive fusion lets the model preserve the efficiency and stability of global context modeling while injecting direct short-range interactions.

\textbf{Gated Rank Selector.}
However, indiscriminately mixing global and local signals may introduce noise. To address this, we introduce a \textbf{Gated Rank Selector}. We define a gating function $\mathcal{G}(\cdot)$ that projects the input context $\boldsymbol{H}_g$ into a modulation latent space to derive a context-aware gate $\boldsymbol{\Gamma}$:
\begin{equation}
    \boldsymbol{\Gamma} = \mathcal{G}(\boldsymbol{H}_g | \boldsymbol{\Phi}) = \sigma(\boldsymbol{H}_g \boldsymbol{W}_{gate} + \boldsymbol{b}_{gate}),
\end{equation}
where $\boldsymbol{\Phi} = \{\boldsymbol{W}_{gate}, \boldsymbol{b}_{gate}\}$ denotes the learnable parameters of the gating projection. This coefficient vector $\boldsymbol{\Gamma}$ acts as a soft switch, dynamically modulating the information flow between the stationary global preferences and volatile local patterns. Consequently, the final modulated output $\boldsymbol{H}_3$ is obtained via a gated residual formulation: 
\begin{equation}
    \begin{aligned}
    \boldsymbol{H}_2 &= \operatorname{RELA}\left(\boldsymbol{H}_g \boldsymbol{W}_{in} + \boldsymbol{b}_{in}\right), \\
     \boldsymbol{H}_3 &=\operatorname{DropPath}\left(\boldsymbol{W}_{out}\left(\boldsymbol{\Gamma} \odot \boldsymbol{H}_2\right) + \boldsymbol{b}_{out}\right) + \boldsymbol{H},
    \end{aligned}
\end{equation}
where $\boldsymbol{H}_2$ incorporates the augmented representation. By removing the linear projection for the value matrix $\boldsymbol{V}$ in the linear attention, we further reduce redundancy. This architecture allows RecGRELA to adaptively shift its focus: relying on linear attention for stable long-term preferences while leveraging the local augmentation branch for transient, bursty interactions.

\subsection{Multi-Layer Perceptron}
To capture the complex features, we further leverage a Multi-Layer Perceptron (MLP) \cite{kang2018self}:
\begin{equation}
    \begin{aligned}
    \boldsymbol{H}_4&=\operatorname{LayerNorm}\left(\boldsymbol{H}_3\right), \\\boldsymbol{H}_L&=\operatorname{Drop}\left(\operatorname{Drop}\left(\operatorname{GELU}\left(\boldsymbol{H}_4\boldsymbol{W}_4+\boldsymbol{b}_4\right)\right)\boldsymbol{W}_5+\boldsymbol{b}_5\right) + \boldsymbol{H}_3,
    \end{aligned}
\end{equation}
where $\boldsymbol{W}_4\in\mathbb{R}^{D\times4D},\boldsymbol{W}_5\in\mathbb{R}^{4D\times D}$ denote the weight matrices, $\boldsymbol{b}_4$, $\boldsymbol{b}_5$ are also the biases. $\boldsymbol{H}_L$ indicates the final output representation, and $\boldsymbol{H}_4$ denotes the output of layer normalization after the GRELA block. To stabilize the training process and enhance gradient flow through the network, we employ a residual connection \cite{he2016deep}. For generality, the subscript (0) indicates that the final user representation is obtained using a single GRELA layer.

\subsection{Prediction Layer and Training}
After the computation of $L$ GRELA blocks and the MLP, we get the final representation of behavior sequences with user interest. Based on this, we compute the dot product between it and the embeddings of all items $\boldsymbol{E} \in \mathbb{R}^{|\mathcal{V}| \times D}$. This result is then scaled using the Softmax function to obtain the interaction probability score:
\begin{equation}
    \hat{\mathcal{Y}}=\operatorname{Softmax}(\boldsymbol{E}^\top\boldsymbol{H}_L).
\end{equation}
Hence, following previous works \cite{kang2018self,liu2024mamba4rec}, we expect to minimize the cross-entropy between the predicted recommendation results $\hat{\mathcal{Y}}$, and the ground truth items $\mathcal{Y}$:
\begin{equation}
    \mathcal{L}_{CE}\left(\mathcal{Y},\hat{\mathcal{Y}}\right)=\mathcal{Y}\log\left(\hat{\mathcal{Y}}\right)+\left(1-\mathcal{Y}\right)\left(1-\log\left(\hat{\mathcal{Y}}\right)\right).
\end{equation}

\subsection{Complexity Analysis}
We analyze the computational complexity of RELA (as shown in Eq.~\eqref{eq:RELA}), focusing on several key components.
To begin with, the kernel function in matrices $\boldsymbol{Q}, \boldsymbol{K}, \in\mathbb{R}^{N\times{d}}$, incurs a computational cost of $\mathcal{O}(Nd)$.
Subsequently, the RoPE operator $\mathcal{R}(\cdot)$ contributes a time complexity that is also $\mathcal{O}(Nd)$.
Furthermore, the matrix multiplication operation has a $\mathcal{O}(Nd^2)$ complexity. This contrasts with the original dot-product attention mechanism, which exhibits a complexity of $\mathcal{O}(N^2d)$.
Consequently, the overall time complexity of the RELA is $\mathcal{O}(Nd^2)$.

\section{Experiments}

\begin{table*}[]
    \centering
    \renewcommand\arraystretch{1.3}
    \caption{Statistics of the experimental datasets.}
    \begin{tabularx}{\textwidth}{@{} *{7}{>{\centering\arraybackslash}X} @{}}
    \toprule
    Datasets & \#Users & \#Items & \#Interactions & Avg. UA & Avg. IA & Sparsity\\ \midrule
    ML-1M     & 6,041 & 3,417  & 999,611        & 165.5 & 292.6 & 95.15\% \\
    ML-32M   & 200,949 & 43,885 & 31,921,467    & 158.9  & 727.4 & 99.63\%\\
    Tmall  & 413,068 & 221,888 & 4,985,558  & 12.1 & 22.5 & 99.99\% \\
    LFM-1B   & 78,113 & 179,515 & 15,628,182    & 200.1  & 87.1 & 99.89\%\\
    Netflix & 120,378 & 16,087 & 21,767,054     & 180.8  & 1,353.1 & 98.86\%\\
    \bottomrule
    \end{tabularx}
    \label{tab:dataset}
\end{table*}

\subsection{Experimental Settings}
\subsubsection{Datasets}
We conduct experiments and verify the effectiveness of our approach on five datasets, including MovieLens-1M and 32M \footnote{https://grouplens.org/datasets/movielens/} (dubbed as ML-1M, ML-32M), Tmall~\footnote{https://tianchi.aliyun.com/dataset/53}, LFM-1B \footnote{https://www.cp.jku.at/datasets/LFM-1b/}, and Netflix \footnote{https://www.kaggle.com/datasets/netflix-inc/netflix-prize-data}. The dataset statistics are shown in Table \ref{tab:dataset}, where \textit{Avg. UA} and \textit{Avg. IA} denote the average actions of users and items. Notably, for MovieLens and Tmall datasets, we filter out users and items with fewer than 5 interactions. For LFM-1B datasets, we use the processed atomic files of LFM1B-Artists (merged) and filter out users with fewer than 50 and more than 500 interactions. Netflix is a huge-scale dataset, containing approximately 100 million interactions and five thousand \textit{Avg. IA}. To simplify the training process, we filter out users with fewer than 50 interactions while retaining items with interactions ranging from 50 to 15,000 for this dataset.

\subsubsection{Baseline Models}
To verify the effectiveness and efficiency of RecGRELA, we make comparisons between RecGRELA and strong RNN-, Attention-, Mamba-based, and frequency-domain-based baselines:
\textbf{SASRec} \cite{kang2018self}: The representative model that uses the self-attention mechanism for sequential recommendation. \textbf{LinRec} \cite{liu2023linrec}: This model reduces the computational costs via linear attention in transformer-based models. We choose SASRec \cite{kang2018self} as the backbone of LinRec. \textbf{LRURec} \cite{yue2024linear}: This method is based on linear recurrent units \cite{orvieto2023resurrecting} for sequential recommendation. \textbf{Mamba4Rec} (Mamba) \cite{liu2024mamba4rec}: This is the first Mamba-based method, which explores applying the Mamba architecture in sequential recommendation tasks. \textbf{EchoMamba4Rec (Echo)} \cite{wang2024echomamba4rec}: This method processes sequences both forward and backward to enhance the sequential representation with Mamba and frequency domain filter. \textbf{EulerFormer (Euler)} \cite{tian2024eulerformer}: This method proposes a transformer variant with complex vectors based on Euler’s formula and a contrastive learning method to improve contextual representations. \textbf{RecBLR} \cite{liu2024behavior}: This model shows a behavior-dependent linear recurrent unit to enhance user behavior modeling with a hardware-aware parallel scan approach. \textbf{SIGMA} \cite{liu2025sigma}: This model employs a flipped Mamba to build a bidirectional framework to improve short sequence representations and contextual modeling. \textbf{MUFFIN}~\cite{muffin}: This model merges global and local filtering blocks to achieve multi-scale pattern recognition.

\begin{table*}[h]
    \centering
    \footnotesize
    \renewcommand\arraystretch{1.3}
    \caption{Performance comparisons on ML-1M, Tmall, ML-32M, LFM-1B, and Netflix datasets. Best and second-best results per row are shown in \textbf{bold} and \underline{underlined} format, respectively. "Improv." denotes the percentage improvement of the best-performing model over the second-best for each metric. The ``$\star$'' denotes the statistical significance ($p$ < 0.05) of the results of RecGRELA compared to the best baseline.}
    \resizebox{1.0\textwidth}{!}{
    \label{tab:overallperfomance}
    \begin{tabular}{llccccccccccc}
    \toprule
    Dataset & Metric & SASRec & LinRec & LRURec & Mamba & Echo & Euler & RecBLR & SIGMA & MUFFIN & RecGRELA & Improv. \\
    \midrule
    \multirow{6}{*}{ML-1M}
     & HR@5   & 0.2045 & 0.2187 & 0.2134 & 0.2211 & 0.2243 & 0.2159 & 0.2293 & \underline{0.2325} & 0.2201 & \textbf{0.2364}$^\star$ & 1.68\% \\
     & NDCG@5   & 0.1384 & 0.1472 & 0.1418 & 0.1521 & 0.1539 & 0.1493 & \underline{0.1604} & 0.1587 & 0.1512 & \textbf{0.1653}$^\star$ & 3.05\% \\
     & MRR@5   & 0.1185 & 0.1263 & 0.1217 & 0.1314 & 0.1328 & 0.1281 & \underline{0.1377} & 0.1345 & 0.1306 & \textbf{0.1419}$^\star$ & 3.05\% \\
     & HR@10  & 0.3012 & 0.3109 & 0.3026 & 0.3230 & 0.3253 & 0.3089 & 0.3258 & \underline{0.3291} & 0.3134 & \textbf{0.3311}$^\star$ & 0.61\% \\
     & NDCG@10  & 0.1746 & 0.1837 & 0.1734 & 0.1869 & 0.1907 & 0.1768 & \underline{0.1917} & 0.1898 & 0.1802 & \textbf{0.1959}$^\star$ & 2.19\% \\
     & MRR@10  & 0.1367 & 0.1448 & 0.1340 & 0.1453 & 0.1496 & 0.1365 & \underline{0.1507} & 0.1472 & 0.1394 & \textbf{0.1545}$^\star$ & 2.52\% \\
    \midrule
    \multirow{6}{*}{ML-32M}
     & HR@5   & 0.1478 & 0.1568 & 0.1513 & 0.1621 & 0.1649 & 0.1587 & 0.1671 & \underline{0.1698} & 0.1602 & \textbf{0.1802}$^\star$ & 6.12\% \\
     & NDCG@5   & 0.0965 & 0.1034 & 0.0989 & 0.1092 & 0.1117 & 0.1056 & 0.1151 & \underline{0.1180} & 0.1068 & \textbf{0.1258}$^\star$ & 6.61\% \\
     & MRR@5   & 0.0812 & 0.0871 & 0.0826 & 0.0923 & 0.0948 & 0.0894 & 0.0980 &  \underline{0.1009} & 0.0906 & \textbf{0.1080}$^\star$ & 7.04\% \\
     & HR@10  & 0.2256 & 0.2397 & 0.2235 & 0.2372 & 0.2401 & 0.2208 & 0.2378 & \underline{0.2426} & 0.2245 & \textbf{0.2519}$^\star$ & 3.83\% \\
     & NDCG@10  & 0.1321 & 0.1402 & 0.1296 & 0.1369 & 0.1398 & 0.1270 & 0.1379 & \underline{0.1423} & 0.1294 & \textbf{0.1490}$^\star$ & 4.71\% \\
     & MRR@10  & 0.1012 & 0.1090 & 0.1015 & 0.1062 & 0.1083 & 0.1003 & 0.1073 & \underline{0.1116} & 0.1021 & \textbf{0.1175}$^\star$ & 5.29\% \\
    \midrule
    \multirow{6}{*}{Tmall}
     & HR@5   & 0.0854 & 0.0895 & 0.0832 & 0.0908 & 0.0921 & 0.0945 & 0.0975 & \underline{0.0982} & 0.0951 & \textbf{0.1022}$^\star$  & 4.07\% \\
     & NDCG@5   & 0.0543 & 0.0599 & 0.0581 & 0.0605 & 0.0618 & 0.0610 & 0.0637 & \underline{0.0645} & 0.0623 & \textbf{0.0675}$^\star$  & 4.65\% \\
     & MRR@5   & 0.0381 & 0.0415 & 0.0401 & 0.0422 & 0.0439 & 0.0495 & 0.0525 & \underline{0.0533} & 0.0503 & \textbf{0.0560}$^\star$  & 5.07\% \\
     & HR@10  & 0.1172 & 0.1245 & 0.1188 & 0.1252 & 0.1234 & 0.1233 & 0.1245 & \underline{0.1281} & 0.1241 & \textbf{0.1335}$^\star$ & 7.23\% \\
     & NDCG@10  & 0.0641 & 0.0692 & 0.0653 & 0.0690 & 0.0685 & 0.0711 & 0.0720 & \underline{0.0724} & 0.0715 & \textbf{0.0776}$^\star$ & 7.18\% \\
     & MRR@10  & 0.0437 & 0.0485 & 0.0441 & 0.0473 & 0.0481 & 0.0456 & 0.0569 & \underline{0.0578} & 0.0468 & \textbf{0.0601}$^\star$ & 3.98\% \\
    \midrule
    \multirow{6}{*}{LFM-1B}
     & HR@5   & 0.0691 & 0.0755 & 0.0732 & 0.0768 & 0.0781 & 0.0745 & 0.0778 & \underline{0.0834} & 0.0754 & \textbf{0.0860}$^\star$  & 3.12\% \\
     & NDCG@5 & 0.0384 & 0.0421 & 0.0401 & 0.0435 & 0.0448 & 0.0410 & 0.0447 & \underline{0.0512} & 0.0419 & \textbf{0.0531}$^\star$  & 3.71\% \\
     & MRR@5  & 0.0279 & 0.0315 & 0.0298 & 0.0322 & 0.0339 & 0.0305 & 0.0339 & \underline{0.0407} & 0.0315 & \textbf{0.0424}$^\star$  & 4.18\% \\
     & HR@10  & 0.1261 & 0.1345 & 0.1288 & 0.1352 & 0.1380 & 0.1333 & 0.1420 & \underline{0.1421} & 0.1348 & \textbf{0.1460}$^\star$ & 2.74\% \\
     & NDCG@10& 0.0576 & 0.0632 & 0.0603 & 0.0640 & 0.0655 & 0.0621 & \underline{0.0702} & 0.0700 & 0.0634 & \textbf{0.0724}$^\star$ & 3.13\% \\
     & MRR@10 & 0.0418 & 0.0465 & 0.0441 & 0.0473 & 0.0481 & 0.0456 & 0.0482 & \underline{0.0484} & 0.0463 & \textbf{0.0502}$^\star$ & 3.72\% \\
    \midrule
    
    \multirow{6}{*}{Netflix}
     & HR@5   & 0.0815 & \underline{0.0900} & 0.0712 & 0.0809 & 0.0827 & 0.0781 & 0.0873 & 0.0897 & 0.0802 & \textbf{0.0922}$^\star$ & 2.44\% \\
     & NDCG@5   & 0.0551 & \underline{0.0606} & 0.0443 & 0.0538 & 0.0559 & 0.0512 & 0.0581 & 0.0604 & 0.0528 & \textbf{0.0623}$^\star$ & 2.81\% \\
     & MRR@5   & 0.0462 & 0.0507 & 0.0369 & 0.0446 & 0.0467 & 0.0431 & 0.0483 & \underline{0.0508} & 0.0446 & \textbf{0.0525}$^\star$ & 3.35\% \\
     & HR@10  & 0.1251 & \underline{0.1368} & 0.1196 & 0.1305 & 0.1327 & 0.1191 & 0.1338 & 0.1359 & 0.1215 & \textbf{0.1386}$^\star$ & 1.32\% \\
     & NDCG@10  & 0.0694 & \underline{0.0765} & 0.0650 & 0.0728 & 0.0736 & 0.0648 & 0.0747 & 0.0763 & 0.0665 & \textbf{0.0792}$^\star$ & 3.53\% \\
     & MRR@10  & 0.0521 & \underline{0.0582} & 0.0484 & 0.0532 & 0.0553 & 0.0495 & 0.0567 & \underline{0.0582} & 0.0511 & \textbf{0.0611}$^\star$ & 4.98\% \\
    \bottomrule
    \end{tabular}
    }
    \label{table:dataset} 
\end{table*}

\begin{figure}
    \centering
    \includegraphics[width=1.0\columnwidth]{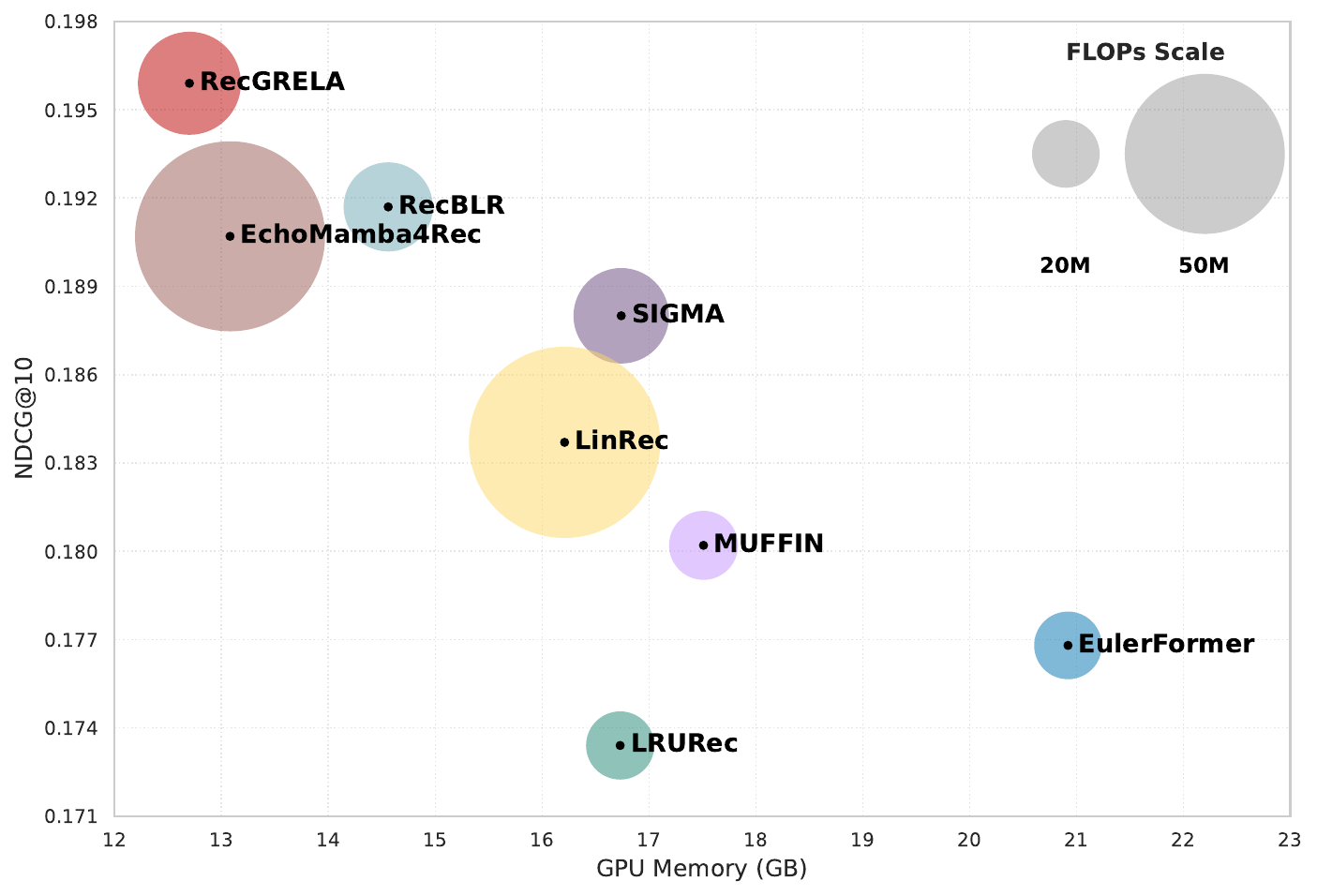}	
    \caption{RecGRELA vs. baselines in terms of the GPU memory and FLOPs of the training stage on the ML-1M dataset.}
    \label{fig:efficiency_all}
\end{figure}

\subsubsection{Evaluation Protocols}
We evaluate performance using three standard metrics: \textit{Hit Rate} (HR), \textit{Normalized Discounted Cumulative Gain} (NDCG), and \textit{Mean Reciprocal Rank} (MRR). These metrics are assessed based on top-$K$ recommendation lists, specifically using cutoff values $K$ = 5 and $K$ = 10, denoted as HR@5, NDCG@5, MRR@5, HR@10, NDCG@10, and MRR@10, respectively.

\subsubsection{Implementation Details}
We utilize the RecBole library \footnote{https://github.com/RUCAIBox/RecBole} \cite{zhao2021recbole,xu2023towards}, a unified framework for implementing sequential recommendations. To ensure a fair comparison, we fix the embedding size to 64, the hidden size of the MLP to 256, and the training batch size to 2048 for all the models. We directly adopt the values reported in their original manuscripts for the remaining hyperparameters of the baseline models. The number of blocks $L$ is set to 4. The number of attention heads $h$ is set to 4.  The kernel size for 1D causal convolution is 4. We set the Dropout and Droppath rate to 0.1. We use Adam \cite{kingma2014adam} with a learning rate of 0.001 for optimizing our model. The maximum sequence length $N$ is set to 200 for all the datasets. The early stopping epochs are 20 for ML-1M and 10 for other datasets. All the experiments are conducted on a single NVIDIA RTX4090. 

\subsection{Overall Performance}
Table \ref{tab:overallperfomance} reports the recommendation performance of RecGRELA compared to seven baselines on five datasets. From the table, we find several observations:

\begin{itemize}
    \item Our RecGRELA outperforms all competing transformer-based, RNN-based, and Mamba-based baselines, with improvements ranging from 0.61\% to 4.98\%. Such a comparison highlights the effectiveness of our design in the long-term sequential recommendation.
    \item The performance improvements are more significant on larger datasets with richer interactions and more average action of items. For example, the Netflix dataset has almost 22M interactions and relatively long item actions (1.3K). Conversely, performance gains are more modest for sparser datasets with fewer interactions, users, and items, such as ML-1M. This phenomenon can be attributed to the data-rich environments providing sufficient information for the model to learn more complex patterns, leading to more accurate recommendation performance.
    \item The improvements in the metrics of NDCG and MRR are typically greater than those in HR. This discrepancy stems from their distinct evaluation criteria: NDCG and MRR employ position-sensitive weighting schemes prioritizing ranking precision, whereas HR operates as a binary relevance indicator. This ranking superiority enables RecGRELA to generate recommendations with enhanced positional accuracy, elevating user-preferred items to top-tier placements while maintaining contextual relevance. This can also be attributed to introducing RoPE, which enhances the model's ability to capture positional information.
\end{itemize}

\subsection{Efficiency Analysis}
We compare the training efficiency of our RecGRELA with state-of-the-art sequential recommendation models in terms of GPU memory and FLOPs. The results are shown in Fig. \ref{fig:efficiency_all}. The results are based on the ML-1M dataset with a sequence length of 200. Our RecGRELA demonstrates superior efficiency and performance, achieving the highest NDCG@10 of 0.1959 with the lowest GPU memory (12.7GB) among the compared models. 

Specifically, our proposed RecGRELA achieves a 2\% performance improvement while requiring only 87$\%$ GPU memory consumption and comparable FLOPs compared to RecBLR. More significantly, when compared with LinRec, our method demonstrates comprehensive superiority across all metrics: it reduces GPU memory usage by 27.6\%, decreases FLOPs by 23.1\%, and achieves a 6.6\% performance lead in NDCG@10. Although EulerFormer and LRURec exhibit relatively low computational FLOPs, they still require many GPU memory while delivering subpar performance. Overall, whether compared to linear attention models (LinRec, EulerFormer), Mamba-based architectures (EchoMamba4Rec, SIGMA), or RNN-based approaches (RecBLR, LRURec), RecGRELA achieves a balance between performance and efficiency.

\begin{table}[t!]
\centering
\caption{Ablation study of RecGRELA on the ML-1M dataset. We report HR@10, NDCG@10, and MRR@10 to demonstrate the individual and combined effects of RELA, ARM, and GRS.}
\label{tab:ablation_components}
\resizebox{0.95\columnwidth}{!}{
  \renewcommand{\arraystretch}{1.4}
  \begin{tabular}{cccccc}
  \toprule
  \textbf{RELA} & \textbf{ARM} & \textbf{GRS} & \textbf{HR@10} & \textbf{NDCG@10} & \textbf{MRR@10} \\
  \midrule
  \XSolidBrush & \XSolidBrush & \XSolidBrush & 0.3109 & 0.1837 & 0.1448 \\
  \Checkmark & \XSolidBrush & \XSolidBrush & 0.3168 & 0.1873 & 0.1479 \\
  \Checkmark & \Checkmark & \XSolidBrush & \underline{0.3216} & \underline{0.1902} & \underline{0.1498} \\
  \midrule
  \Checkmark & \Checkmark & \Checkmark & \textbf{0.3311} & \textbf{0.1959} & \textbf{0.1545} \\
  \bottomrule
  \end{tabular}
}
\end{table}

\begin{table*}[!t]
    \centering
    \renewcommand\arraystretch{1.3}
    \caption{Sensitivity analysis on different numbers of layers. The sequence length is set to 200. Time indicates the training time per epoch, and Memory indicates the GPU memory.}
    \label{table4}
    \begin{tabularx}{\textwidth}{@{}
                                c
                                *{8}{>{\centering\arraybackslash}X}
                                @{}}
    \toprule
    \#Layers & HR@5 & NDCG@5 & MRR@5 & HR@10 & NDCG@10 & MRR@10 & Time & Memory \\
    \midrule
    1 & 0.1512 & 0.0987 & 0.0815 & 0.2874 & 0.1669 & 0.1301 & 20.70s & 5.96GB \\
    2 & 0.1699 & 0.1050 & 0.0899 & 0.3184 & 0.1838 & 0.1426 & 39.49s & 7.71GB \\
    3 & 0.1753 & 0.1088 & 0.0921 & 0.3228 & 0.1866 & 0.1450 & 58.00s & 10.16GB \\
    4 & \textbf{0.1865} & \textbf{0.1152} & \textbf{0.0988} & \textbf{0.3311} & \textbf{0.1959} & \textbf{0.1545} & 76.50s & 12.70GB \\
    5 & 0.1801 & 0.1115 & 0.0954 & 0.3280 & 0.1916 & 0.1499 & 94.88s & 15.14GB \\
    6 & 0.1776 & 0.1109 & 0.0948 & 0.3242 & 0.1912 & 0.1506 & 113.43s & 17.68GB \\
    \bottomrule
    \end{tabularx}
    \label{tab:sensitivity_layers}
\end{table*}


\subsection{Ablation Study}
\subsubsection{Core Components}
To investigate how the core components contribute to our design, we conduct an ablation study to evaluate the individual and combined effects of the Rotary-Enhanced Linear Attention (RELA), Adaptive Rank Modulator (ARM), and Gated Rank Selector (GRS). We use a basic linear attention module as the baseline and incrementally add the proposed components to evaluate their impact on the ML-1M dataset. Table \ref{tab:ablation_components} reports the evaluation on HR@10, NDCG@10, and MRR@10. From this table, we have the following observations: 
\begin{itemize}
    \item Replacing traditional self-attention with the RELA provides a noticeable performance gain (NDCG@10 improves from 0.1837 to 0.1873). This demonstrates that explicitly incorporating relative positional encoding via RoPE enhances the model's ability to capture sequential dependencies more effectively than standard formulations.
    \item The subsequent addition of the ARM further elevates performance. This highlights the necessity of explicitly modeling dense local user interactions, addressing the low-rank deficiency inherent in linear attention, and injecting crucial short-term preference signals.
    \item Integrating the GRS to form the complete RecGRELA model yields the best performance. This confirms the vital role of the SiLU-based gating mechanism in dynamically modulating and harmonizing stable long-term user preferences with transient local shifts.
\end{itemize}

\begin{figure}
    \centering
    \includegraphics[width=1.0\columnwidth]{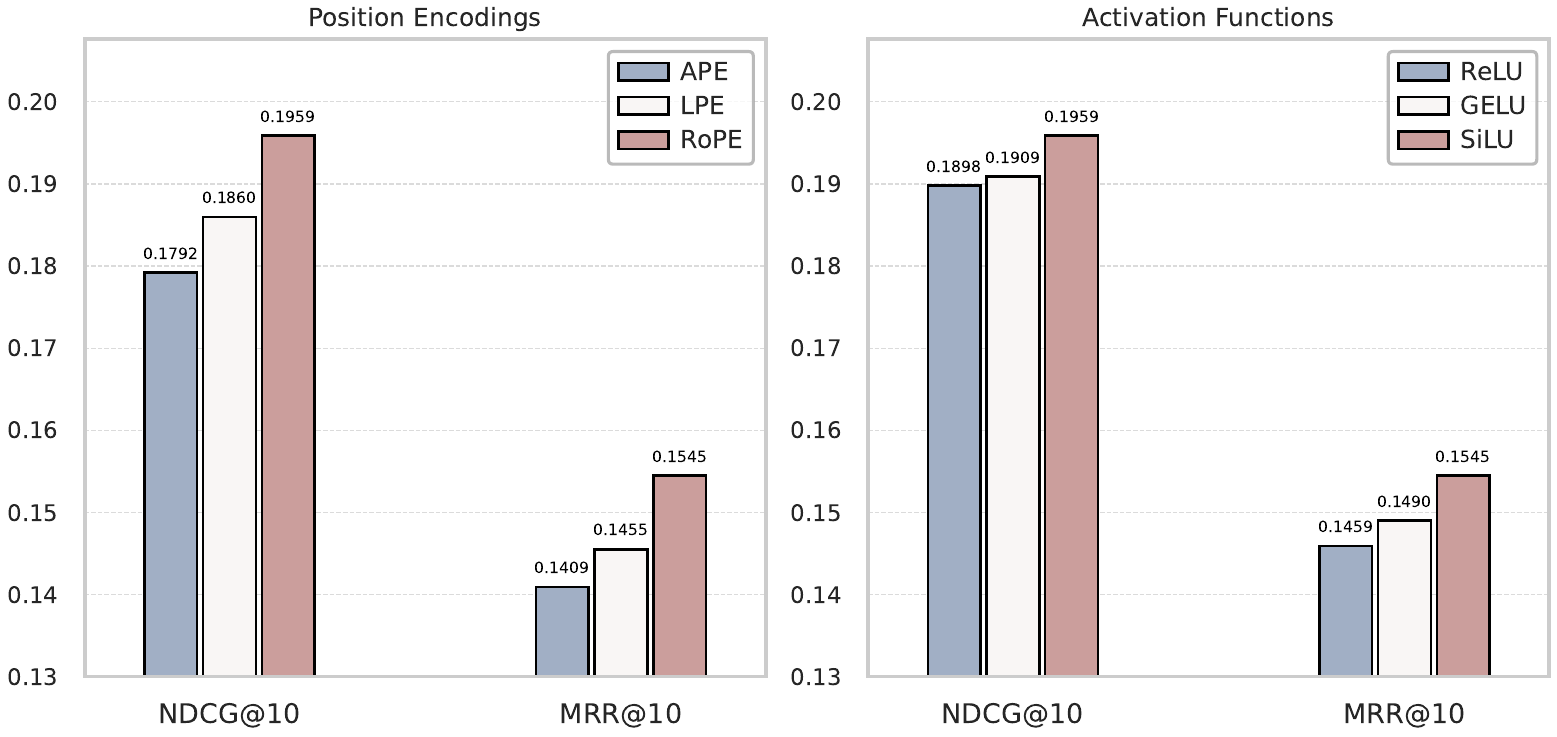}	
    \caption{The performance of different position encodings and activation functions.}
    \label{fig:act_pos}
\end{figure}

\subsubsection{Position Encoding and Activation Function}
We further investigate the impact of position encoding strategies on RecGRELA. As shown in Fig. \ref{fig:act_pos}, the proposed RoPE integration demonstrates superior performance across all metrics while achieving faster training speeds and lower memory consumption than alternative position encoding methods. These empirical results validate the effectiveness of RoPE in enhancing representational capacity within the RecGRELA framework.

We also evaluate the impact of different activation functions on the model's performance. We replace the function in the gated mechanism and the MLP. As shown in Fig. \ref{fig:act_pos}, the SiLU activation function outperforms ReLU and GELU across all metrics, demonstrating the superior performance of SiLU in capturing complex user behavior and enhancing the model's representational capacity.

\begin{figure}
    \centering
    \includegraphics[width=1.0\columnwidth]{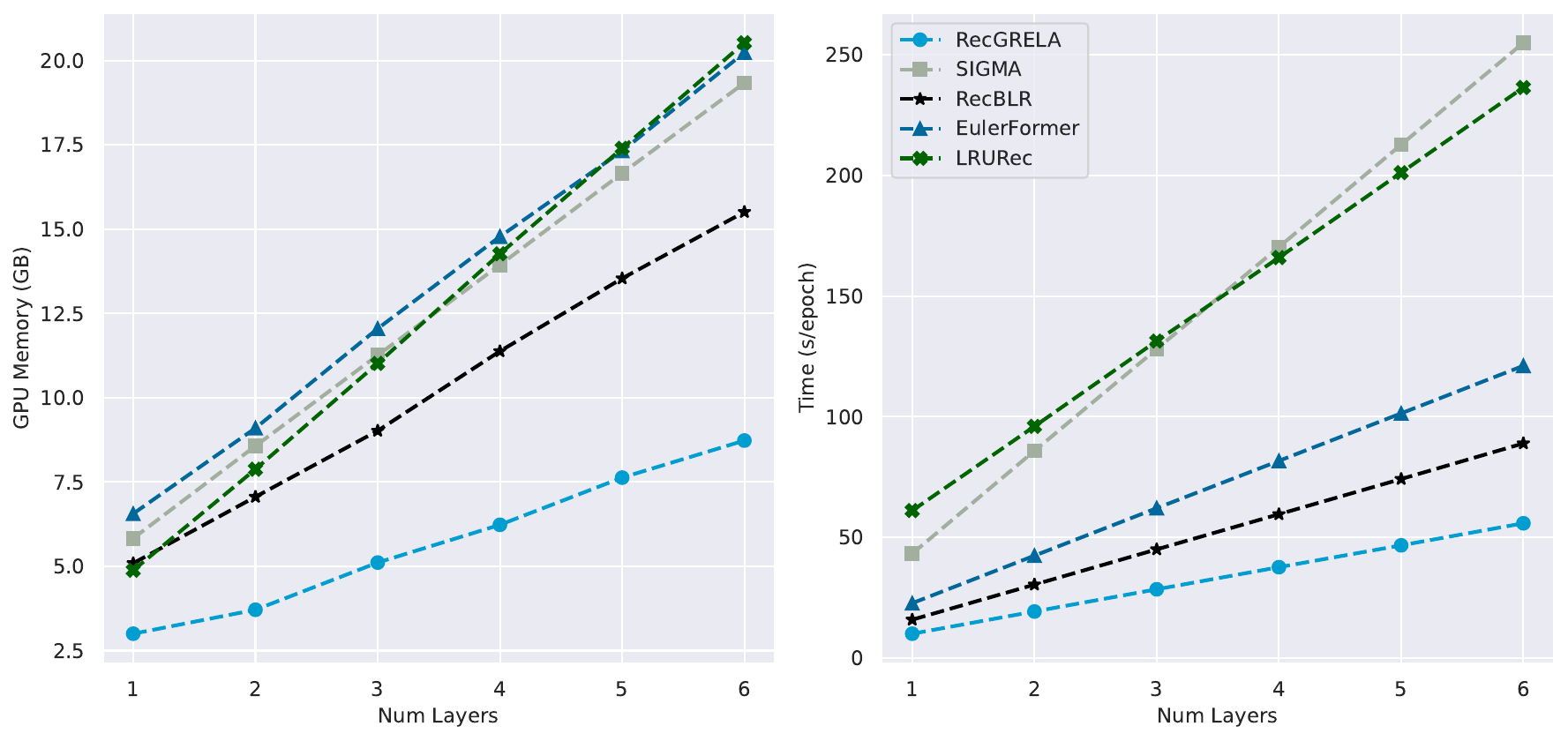}	
    \caption{Impact of different stacked layers on computational costs. We set the sequence length to 100 to avoid GPU memory errors.}
    \label{fig:efficient_hyper}
\end{figure}

\subsection{Sensitivity Analysis}
In this section, we conduct experiments on the \textbf{ML-1M} dataset to analyze the influence of three hyperparameters: the sequence length $N$, the number of stacked GRELA blocks $L$, the number of attention heads $h$, and the Dropout rates.

\subsubsection{The Number of Layers}
The results in Table \ref{tab:sensitivity_layers} reveal that stacking more RecGRELA layers fails to consistently enhance recommendation quality while markedly degrading computational efficiency, primarily due to the overfitting problem of multiple RecGRELA layers. Specifically, model performance peaks at four stacked layers, beyond which further depth expansion leads to metric degradation with substantial computational overhead. Notably, the three-layer configuration achieves competitive performance with state-of-the-art baselines, demonstrating the architecture's effectiveness in balancing representational capacity and operational efficiency. Furthermore, our comparative analysis of layer-wise efficiency between RecGRELA and SIGMA reveals distinct scaling patterns. As illustrated in Fig. \ref{fig:efficient_hyper}, the growth trend of computational occupation of RecGRELA is relatively gradual, whereas that of SIGMA shows a steeper increase, indicating that RecGRELA exhibits better scaling capability when the training requires more layers.

\begin{figure}
    \centering
    \includegraphics[width=1.0\columnwidth]{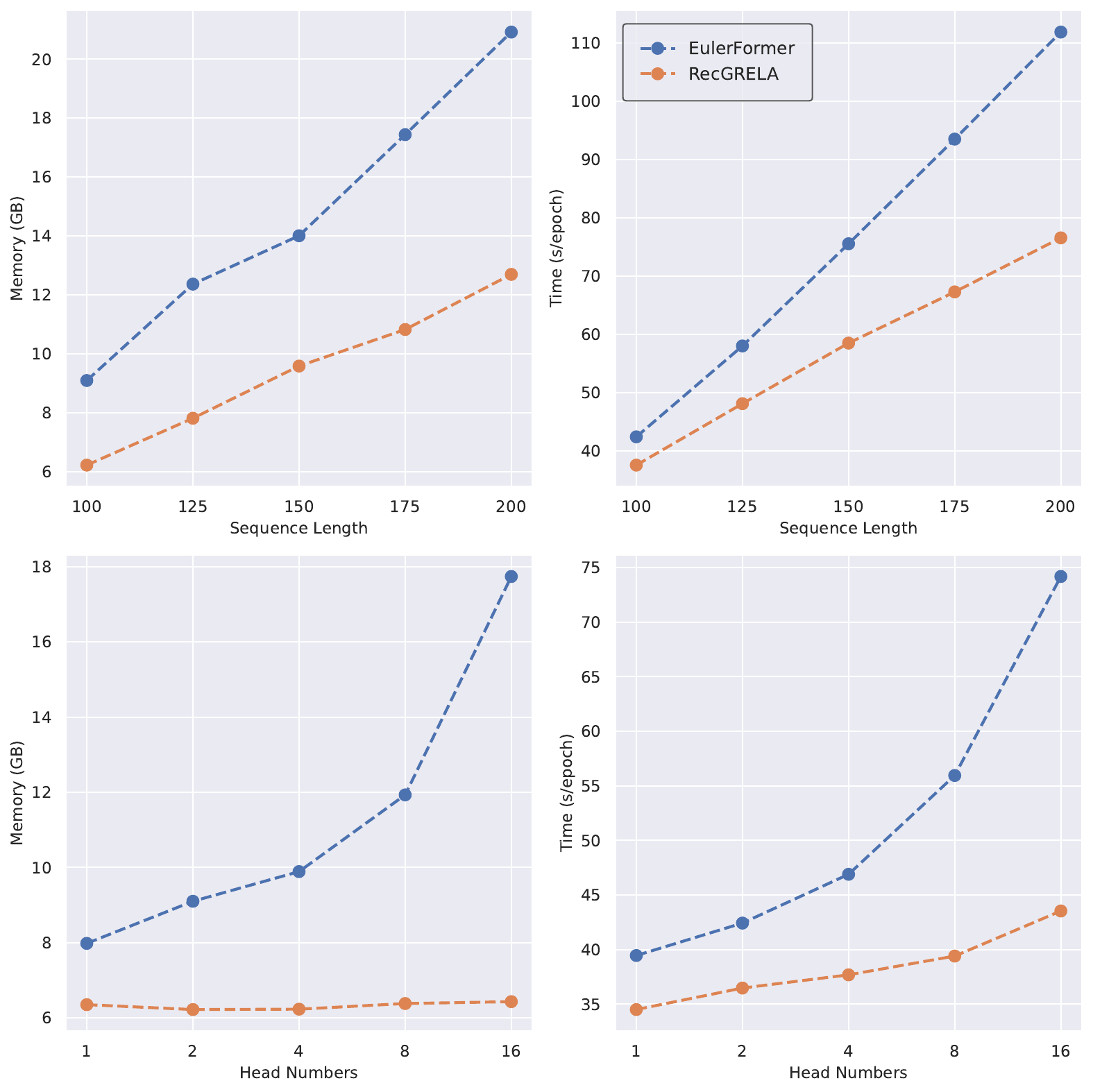}	
    \caption{Impact of sequence length and attention head variations on computational costs. When investigating the impact of sequence length, we set the attention head to 2 for EulerFormer and 4 for RecGRELA. When investigating the impact of attention heads, we set the sequence length to 100 to avoid GPU memory errors. Other hyperparameters are the same as the main experiments.}
    \label{fig:time_gpu_with_euler}
\end{figure}

\subsubsection{The Number of Sequence Length}
To examine training cost variations across sequence lengths in long-term SRSs, we evaluate maximum sequence lengths from 100 to 200 with 25-step increments. Fig. \ref{fig:time_gpu_with_euler} demonstrates that RecGRELA maintains consistently lower computational complexity than EulerFormer across all tested configurations.

To evaluate the model's performance characteristics with varying input lengths, we varied the maximum sequence length processed by the model, testing values from 100 to 200, while keeping other hyperparameters constant. The impact on recommendation performance was measured using NDCG@10 and MRR@10, as illustrated in Fig. \ref{fig:hyper_0507}. The green and blue lines show the model's performance exhibits relative stability for sequence lengths between 100 and 150. Notably, performance significantly improves when the sequence length increases to 200. This suggests that while the model performs consistently well with shorter sequences, capturing longer-range dependencies present in sequences of length 200 yields a clear benefit in terms of recommendation performance.

\subsubsection{The Number of Attention Heads}
We evaluate the impact of the number of attention heads on model efficiency. As shown in Fig. \ref{fig:time_gpu_with_euler}, the results demonstrate that RecGRELA maintains stable GPU memory and training time across varying head counts, while EulerFormer exhibits proportional increases in computational overhead with expanded attention heads. RecGRELA applies rotary position encoding directly to the full embedding dimension, avoiding repetitive computations across attention heads as EulerFormer does. Our method proves that handling position encodings at the embedding level saves computation compared to per-head processing.

We further investigated the model's performance with respect to the number of attention heads. The results are also shown in Fig. \ref{fig:hyper_0507}. Both metrics consistently demonstrate that using four attention heads yields the best performance for our model configuration on this dataset. Utilizing fewer heads (two) appears insufficient to capture the complexity of user-item interactions while employing significantly more heads does not offer further benefits and may even slightly degrade performance. This could be attributed to potential redundancy or fragmentation of attention patterns with excessive heads. Therefore, four attention heads represent the optimal setting identified in our experiments.

\subsubsection{The Dropout and Droppath Rates}
\begin{figure}
	\centering
	\includegraphics[width=1.0\columnwidth]{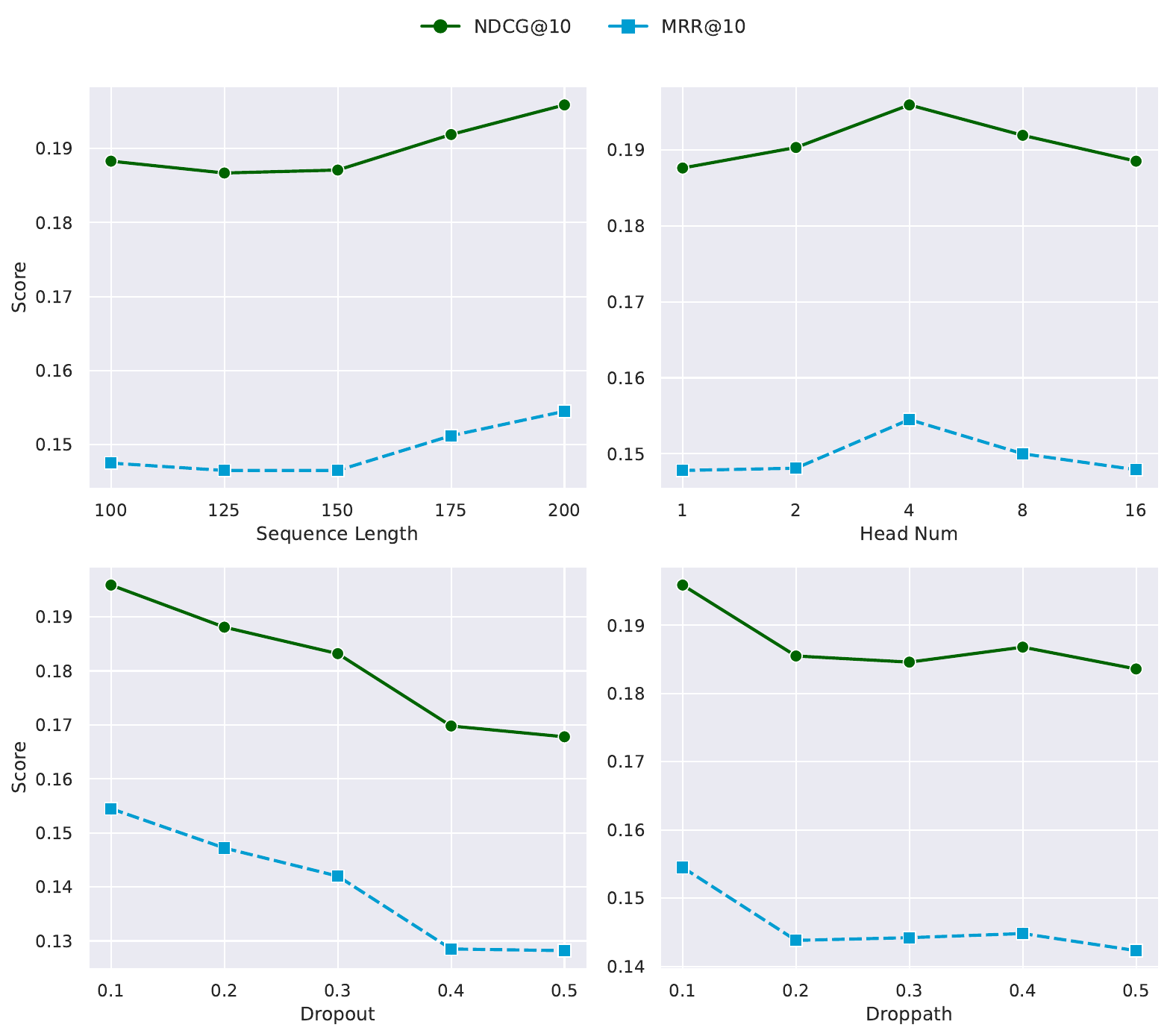}	
	\caption{Influence of the sequence length, head numbers, Dropout, and Droppath on various metrics.}
	\label{fig:hyper_0507}
\end{figure}

Regularization techniques like Dropout and Droppath prevent overfitting and improve model generalization. We assessed the model's sensitivity to the rates of these two techniques, varying each from 0.1 to 0.5. Fig. \ref{fig:hyper_0507} displays the impact on NDCG@10 and MRR@10. The sensitivity analysis suggests that lower regularization rates (0.1 for both) are preferable. Higher rates, particularly for Dropout, lead to substantial performance degradation. This highlights the importance of carefully tuning regularization hyperparameters, as overly aggressive regularization can harm performance more than it helps prevent overfitting in sequential recommendation.

\begin{figure*}
    \centering
    \includegraphics[width=0.95\textwidth]{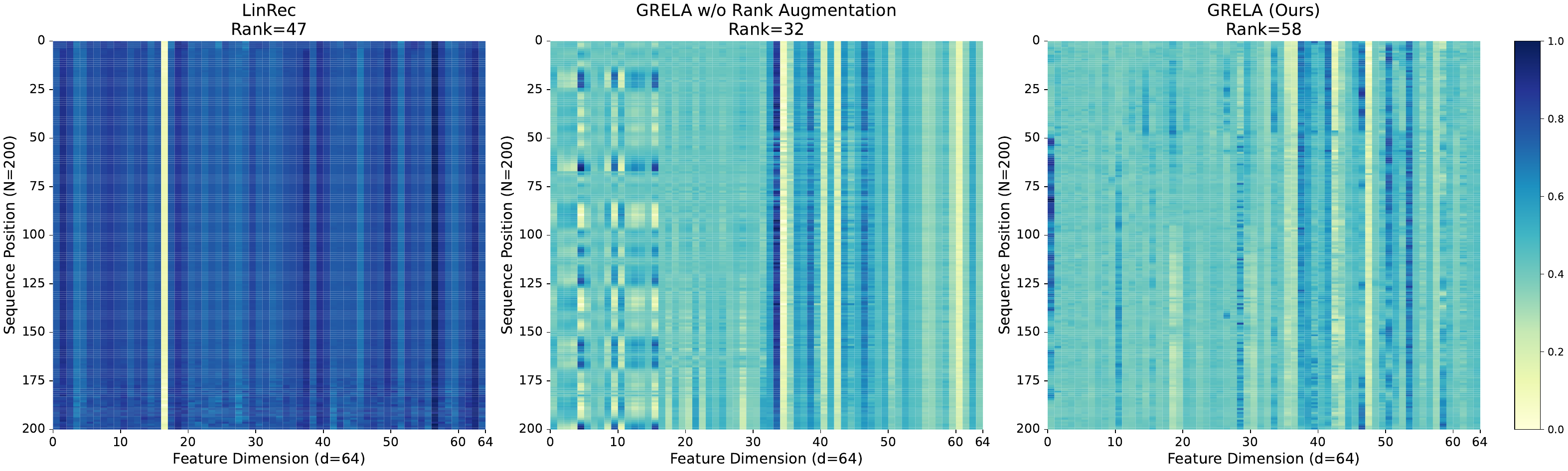}	
    \caption{Heatmap of output feature maps from the last layer.}
    \label{fig:attmap}
\end{figure*}

\subsection{Visualization Analysis}

To further investigate the impact of the proposed Rank Augmentation Branch on the representational capacity of our model, particularly regarding the theoretical motivation to increase feature matrix rank, we conduct a visualization analysis on the \textbf{ML-1M} dataset. We visualize the output feature maps from the last layer for three distinct models: first, LinRec, which serves as a standard linear attention baseline; second, GRELA without the Rank Augmentation Branch, representing our architecture with this specific component ablated; and third, the complete GRELA model, labeled as Ours. The resulting heatmaps display the feature matrix structure for a sequence length $N$ of 200 and a feature dimension $d$ of 64. We also compute the rank for each matrix. Given the feature dimension $d$ = 64, this is the maximum possible rank in this configuration.

Fig. \ref{fig:attmap} helps show why our design is effective by visualizing the resulting feature matrices. The heatmap on the left (LinRec), shows a feature matrix with a rank of 47. The middle heatmap is for GRELA without the Rank Augmentation Branch, yielding a feature matrix with the lowest rank at just 32. Strong horizontal lines and repeating patterns indicate that the features are too similar and lack local detail. This low matrix rank confirms reduced expressive power and information content. In contrast, the heatmap from our full GRELA model with the Rank Augmentation Branch, displays a much more complex and detailed feature matrix. The strong lines disappear, showing more varied features. Significantly, its matrix rank increases to 58, closely approaching the theoretical maximum rank of 64.

The Rank Augmentation Branch, which uses causal convolution, adds important local details to the features. This helps fix the low-matrix rank problem that happens with just linear attention. Adding these local details makes the overall matrix rank higher. As a result, the features become richer and show more variety. We can see this improvement in the heatmap, and the higher matrix rank confirms it. This proves the Rank Augmentation Branch works well to get around the limits of linear attention. It helps the model capture immediate local preferences and crucial long-term dependencies within the user's interaction sequence.

\section{Related Work}
\subsection{Sequential Recommendation}
Earlier fundamental models of sequential recommendation focus on using Convolutional Neural Networks \cite{tang2018personalized,10.1145/3397271.3401156} and Recurrent Neural Networks (RNNs) \cite{hidasi2015session,li2017neural}.
Due to RNNs' performance limits and poor parallel processing, recent sequential recommendation studies have increasingly turned to Transformer-based architectures \cite{wu2020sse,li2021lightweight,du2023frequency,fan2022sequential,zhang2023adaptive,zhao2023user,li2023strec} or MLPs \cite{long2024learning,zhou2022filter,li2023automlp} for better efficiency and scalability. The most representative work is SASRec \cite{kang2018self}, which is the first to use the self-attention mechanism for capturing user interest features from interaction sequences. BERT4Rec \cite{sun2019bert4rec} adopts a deep encoder-based bidirectional model with self-attention to model user behavior sequences, achieving strong recommendation performance. However, the Transformer architecture suffers from quadratic computational complexity with respect to sequence length due to the dot-product attention operation, which limits its efficiency for sequence modeling tasks. Recently, many models with linear computational complexity have emerged as alternatives to the original Transformer architecture, such as linear attention \cite{liu2023linrec,tian2024eulerformer}, linear RNNs \cite{liu2024behavior, yue2024linear}, and state space models \cite{liu2024mamba4rec, qu2024ssd4rec,fan2024tim4rec}. While LinRec \cite{liu2023linrec} adopts linear attention to reduce the computational complexity of the dot-product attention, they still rely on learnable position encodings to capture positional information. UNET4Rec~\cite{wang2024uet4rec} propose a U-net Transformer that utilizes a 1D convolutional encoder to heavily compress sequence embeddings prior to the Transformer block. EulerFormer \cite{tian2024eulerformer} introduces a novel Euler transformation with a rotary form for unified modeling of both semantic and positional difference, yet uses the dot-product attention, which results in computational costs comparable to LinRec. To overcome these specific limitations, RecGRELA combines linear attention with rotary position encodings and introduces a gated mechanism to model both local and long-term user interests.

\subsection{Linear Attention}
As opposed to dot-product attention, linear attention \cite{katharopoulos2020transformers} replaces the Softmax with separate kernel functions, thereby reducing computational complexity to $\mathcal{O}(N)$ through a change in computation order. Though efficient, designing a linear attention module as effective as softmax attention is a nontrivial problem. Although linear attention has achieved some success in other domains, such as computer vision \cite{shen2021efficient,qin2022cosformer,han2024demystify,han2024agent} and neural language processing \cite{wang2020linformer,qin2024hierarchically, yang2023gated}. In sequential recommendation, despite existing linear attention-based methods \cite{fan2021lighter,liu2023linrec} decreasing the computational complexity, they cannot effectively model position encoding within behavior sequence and fail to capture local preference in long-term interactions. This motivates us to design a more efficient and effective linear attention-based module to address these issues.  In response, we present RecGRELA, which aims to realize the full potential of linear attention in sequential recommendation by overcoming these critical challenges in positional awareness and local preference modeling.

\section{Conclusion and Future Work}
In this work, we propose RecGRELA, a new paradigm for long-term sequential recommendation. By integrating rotary position encoding into a linear attention mechanism, our Rotary-Enhanced Linear Attention module captures long-range dependencies with reduced computational complexity. To model local interaction in long-range dependencies, we design an Adaptive Rank Modulator and provide theoretical insight from the perspective of the rank of the matrix. To address the difficulty of the model in perceiving and adapting to dynamic user preference shifts, we propose a Gated Rank Selector that dynamically balances local and long-term representations. Extensive experiments on five benchmark datasets show that RecGRELA outperforms state-of-the-art methods based on RNNs, Transformers, and Mamba architectures, achieving superior performance with significantly lower memory overhead.

In the future, we plan to explore the application of RecGRELA in other recommendation scenarios, such as session-based and multi-modal recommendations. Since linear attention shows great potential in various domains, we also plan to explore more efficient attention mechanisms for sequential recommendation tasks.


\printcredits

\bibliographystyle{cas-model2-names}

\bibliography{cas-refs}



\end{document}